# Exciton-Polaritons in Hybrid Inorganic-organic Perovskite Fabry-Pérot Microcavity

*Shuai Zhang,[†,‡,§] Qiuyu Shang,[ſ] Wenna Du,[†] Jia Shi,[†,‡] Yang Mi,[†] Jie Chen,[†,‡] Zhiyong Wu,[†] Yuanzheng Li,[†] Mei Liu,[§, ℓ*] Qing Zhang,[ſ, £*] and Xinfeng Liu[†,*]*

[†]Division of Nanophotonics, CAS Key Laboratory of Standardization and Measurement for Nanotechnology, CAS Center for Excellence in Nanoscience, National Center for Nanoscience and Technology, Beijing 100190, P. R. China

[‡]University of Chinese Academy of Sciences, 19 A Yuquan Rd, Shijingshan District, Beijing 100049, P. R. China

[§]School of Physics and Electronics, Shandong Normal University, Jinan 250014, P. R. China

[ſ]Department of Materials Science and Engineering, College of Engineering, Peking University, Beijing 100871, P. R. China

[ℓ]Institute of Materials and Clean Energy, Shandong Normal University, Jinan, 250014, P.R. China

[£]Research Center for Wide Gap Semiconductor, Peking University, Beijing 100871, P. R. China

[*]Email address: liuxf@nanoctr.cn, Q_zhang@pku.edu.cn, liumei@sdnu.edu.cn

**ABSTRACT**

Exciton-polaritons in semiconductor microcavities generate fascinating effects such as long-range spatial coherence and Bose-Einstein Condensation (BEC), which are attractive for their potential use in low threshold lasers, vortices and slowing light, *etc*. However, currently most of exciton-polariton effects either occur at cryogenic temperature or rely on expensive cavity fabrication procedures. Further exploring new semiconductor microcavities with stronger exciton photon interaction strength is extensively needed. Herein, we demonstrate room temperature photon exciton strong coupling in hybrid inorganic-organic $CH_3NH_3PbBr_3$ Fabry-Pérot microcavities for the first time. The vacuum Rabi splitting energy is up to ~390 meV, which is ascribed to large oscillator strength and photon confinement in reduced dimension of optical microcavities. With increasing pumping energy, exciton-photon coupling strength is weakened due to carrier screening effect, leading to occurrence of photonic lasing instead of polartion lasing. The demonstrated strong coupling between photons and excitons in perovskite microcavities would be helpful for development of high performance polariton-based incoherent and coherent light sources, nonlinear optics, and slow light applications.

**Introduction**

Polaritons are bosonic quasiparticles occurring under strong light matter interaction conditions. That is, electric dipoles under the excitation of light can strongly interact with photons. Different type of polaritons can be formed through the coupling of photons with optical phonons, excitons, plasmons, *etc*.[1] For semiconductors, polaritons are often associated with bound electron-hole pairs called excitons, *i.e.*, exciton-polariton. Due to the hybrid nature, polaritons exhibit unique properties derived from their original components. One distinctive property is their ultra-small effective mass compared to that of an electron ($10^{-4}$~$10^{-5}$ times),[2] which can result in large de Broglie wavelength and therefore make it easier to produce Bose-Einstein condensates (BECs)[3]. Indeed, room temperature polariton condensates have been observed in wide bandgap semiconductors and organic molecules based microcavities[2]. Other polariton induced phenomena, such as long-range spatial coherence[4], superfluidity[5], or vortices[6] have also been observed. These fascinating properties make exciton-polariton practical interest to technological devices like low-threshold lasers[7], optical switching[8], and slow light applications[9].

To realize exciton polariton, strong coupling between exciton and cavity photons needs to overcome dissipative rate of excitons and photons. The coupling strength is expressed as vacuum Rabi splitting energy $\Omega=2\hbar g$, where $g$ is the vacuum Rabi frequency $g \propto \sqrt{n(V)*f(V)/V_m(V)}$, with $n(V)$ the number of exciton oscillators, $f(V)$ the oscillator strength, $V_m(V)$ the mode volume. Therefore, aiming to promote exciton-photon coupling strength, reduced dimensional semiconductor structure, *i.e.* quantum wells (QW) or quantum dots (QD) are adopted to enhance exciton oscillator strengths. On the other hand, optical cavities with high quality (*Q*) factor are adopted to increase electromagnetic field density and therefore enlarge photon lifetime[7, 10]. For example, the first observation of cavity polariton[11] is demonstrated in a GaAs QW embedded between distributed Bragg mirrors, which however is realized under 20 K due to small exciton binding energy. Further, through using wideband gap inorganic semiconductors such as GaN[12] and ZnO[13] NWs with large intrinsic exciton binding energies of 30 meV[12] and 60 meV[14], larger Rabi splitting energy even BEC is realized at room temperature. However, the achievement of inorganic semiconductors based cavity polaritons strongly rely on expensive and complicated fabrication proceedings. As a contrast, organic semiconductors are ease of synthesis and show large oscillator strength due to the small Bohr radii of excitons[2]. This can be advantageous for the relaxation of polaritonic state since high-energy optical phonons predominate the relaxation[15]. However, the weak nonlinearities of Frenkel exciton[4], together with large threshold density, restrict the occurrence of exciton-polariton condensation in organic semiconductors based exciton-polaritons. Hybrid organic-inorganic materials, combining the advantages of both organic and inorganic materials, usually have large exciton binding energies, solution processability and good crystalline quality[2]. The excitons of Wannier−Mott type lead to stronger nonlinearities below the Mott transition density.[2] For example, anticrossing of 190 meV was observed in spin-coated metal-halide perovskites $(C_6H_5C_2H_4-NH_3)_2PbI_4$ thin film without high-*Q* cavity at room temperature[16,17]. In contrast, organic metal-halide perovskites $CH_3NH_3PbX_3$ (X= Cl, Br, I) with smaller organic cations are ease to form high quality single crystals with low defect density, thus advantageous for high $Q$ factor optical cavities[18,19].

In this work, exciton-polariton effect in $CH_3NH_3PbBr_3$ (MAPbBr$_3$) nanowires (NWs) is studied by spatially resolved photoluminescence (PL) spectroscopy. With decreasing of NW dimension, Rabi splitting energy increases from 268 meV (for bulk polariton) towards 390 meV for microcavity polaritons, respectively. Experimental and theoretical studies reveal larger Rabi splitting energy in small NWs, which is attributed to increase of oscillator strength and decrease of photonic mode volume. With increasing of excitation fluence, exciton-photon coupling strength is weakened by decrease of oscillator strength owing to carrier screening effect, which then hinders the occurrence of polariton lasing in MAPbBr$_3$ Fabry-Pérot cavities.

**Results and discussion**

MAPbBr$_3$ NWs were synthesized by a well-established one-step solution method (*Supporting information*, **Note S1**).[20] Equimolar quantities of Methylammonium bromide (MABr) and PbBr$_2$ were dissolved in N, N-Dimethylformamide (DMF), yielding stock solution of MAPbBr$_3$. 15 μL of as-prepared MAPbBr$_3$ solution was dipped onto ITO glass substrate for a day under $CH_2Cl_2$ atmosphere at room temperature. Optical image shows that large amount of perovskite NWs as well as platelets are grown on substrate (**Figure 1a**). Scanning electron microscopy (SEM,

**Figure 1b**) measurements show that these NWs have rectangle cross section, suggesting cubic perovskite crystalline structure. The length of perovskite NWs ranges from 2 to 30 μm; the width and height range from 200 nm to 1 μm (*Supporting Information*, **Figure S5**). **Figure 1c** shows a typical AFM image suggests root mean average roughness of 3.4 nm for as-grown perovskite NWs. Transmission electron microscope (TEM, **Figure 1d**) and X-ray diffraction (XRD, **Figure 1e**) were performed to investigate crystal structure of as prepared perovskite NWs. TEM image of a single $MAPbBr_3$ NW and corresponding selected area electron diffraction (SAED) along [100] zone axis reveal that as-grown $MAPbBr_3$ NWs are single crystalline with cubic phase[21]. The XRD pattern contains a set of strong (00l) diffraction peaks, indicating (001) series of crystal planes facets exposed, which are consistent with the SEAD results. Diffraction peaks from high-index lattice planes suggests high crystalline quality, indicating a good match with previous report of perovskite micro crystals.[22] Optical absorption and PL spectra are shown in **Figure 1f**. Clear excitonic absorption peak at 528 nm (2.35 eV) beyond broad absorption edge implies a relatively large exciton binding energy. Further, room temperature PL spectra suggests that as prepared $MAPbBr_3$ shows single excitonic emission at 545 nm (2.28 eV) with full width at half-maximum (FWHM) of ~ 9.74 nm (41meV). The FWHM of PL spectra, thereby damping energy, extracted out from PL is much smaller than of $MAPbBr_3$ quantum dots[23] (21 nm, ~96 meV) and film [24] ( 23 nm, ~96 meV), suggesting high crystalline quality of as-grown samples. Moreover, the large exciton binding energy and low damping energy is promising for realizing strong

exciton-photon coupling.

MAPbBr$_3$ NWs naturally form Fabry-Pérot optical cavities with two end-facets severing as mirrors. Light is confined inside the NWs and oscillates between two end-facets, which can interact with excitons of semiconductor, here MAPbBr$_3$. A home-built spatially resolved PL spectroscopy of perovskite microcavities is conducted to investigate exciton polariton effect in as-grown MAPbBr$_3$ microcavities. As shown in **Figure 2a**, continuous wave 405 nm laser is focused by 100× objective to excite one end of perovskite NW. Then PL is generated and propagates along the NW and the emission from the other end is collected by the same objective in the reflective configuration. **Figure 2b** shows PL image of individual NW excited by 405 nm continuous wave (CW) laser at left end (P$_1$), middle (P$_2$) and right end (P$_3$) of the NW. PL emission from excited position and NW end can be resolved, which suggests good photon propagation and confinement inside the NW. PL spectra of left end exhibits significant redshifts from 2.3 eV to 2.24 eV as the excitation position moves from P$_1$ to P$_3$ mainly due to self-absorption during PL propagation (**Figure 2c**). Moreover, sharp peaks with FWHM of 2.04-3.87 nm can be resolved above spontaneous emission background. The spacing between two adjacent peaks is proportional to the inverse of NW length, which suggests that these peaks are corresponding to longitudinal photonic modes oscillated between two end-facets of the NWs (Fabry-Pérot Cavity, *Supporting information*, **Note S2)**. PL spectra collected from one end of an individual NW (**Figure 2d)** is further deconvoluted to probe microcavity photonic modes. The PL spectra can be well fitted by multi-Lorentzian functions due to a series of Fabry-Pérot (F-P) cavity modes. With the increasing of photon energy, the mode spacing between two adjacent modes decreased gradually, which could not be explained by pure photonic modes. The significant dispersion of cavity photons is similar as that observed in CdS microcavity[25], which is demonstrated by the formation of exciton polarities.

Energy (*E*) *versus* wavevector (*k*) dispersion is studied to probe exciton-polariton effects in as-grown MAPbBr$_3$ NWs. In wavevector space, F-P cavity resonant modes can be placed with integral multiples of $\pi/L_Z$ ($L_Z$, wire length). Therefore energy-wavevector dispersion can be determined by plotting mode energy versus incremental multiples of $\pi/L_Z$. **Figure 3a-c** shows *E-k* dispersion of MAPbBr$_3$ NWs with width of 0.28, 0.27 and 0.32 μm and length of 8.22, 5.8 and 3.66 μm respectively. The corresponding spatial resolved micro-photoluminescence (PL) is shown in **Figure S6**. The dispersion curves can be explained by simulation of rectangular cross-section waveguide using finite element method (FEM). To include the effect of exciton-polariton, dielectric functions $\varepsilon(\omega)$ in the *E-k* dispersion relation of NWs $E(\omega, k) = \hbar\omega = \hbar ck/\sqrt{\varepsilon(\omega)}$ is introduced as the coupled oscillator model. As shown in **Figure 3**, the experimental data could be well fitted by the two-coupled oscillator model. The simulation condition, parameters and electric field distribution of different waveguide modes are available in supporting information **Note S3** and **Note S4**. Clear anticrossing behavior is observed according to experimental and simulation results, suggesting the occurrence of strong exciton photon coupling. For the NW with width × length of 0.28×8.22 μm, a fit was obtained by strong coupling of exciton to fundamental photonic mode using the longitudinal-transverse splitting (L-T splitting energy, $\Delta E_{LT}$) of 11 meV, a quantity which is proportional to the oscillator strength.[10] The strong exciton photon coupling strength is highly dependent on NW size. When NW size decreases from 0.27×5.8 μm to 0.32×3.66 μm, L-T splitting energy increases from 25 meV to 33 meV, as shown in **Figure 3b-c**. Meanwhile, coupling strength, evaluated via Rabi splitting energy, as is indicated by the smallest energy difference between upper and lower polariton branch, also increases as the NW size decrease. Enhanced coupling strength in smaller NWs manifest as strong curvature of dispersion curve, indicating that group index increases significantly near the resonant energy (see **Figure 3d**). With increasing of coupling strength, the group index grows even faster in high energy region. The maximum group index enhancement is up to 2.3 times compared to the value of smallest coupling strength. This large altering in group index can be advantageous for slow light applications such as optical buffer, optical fiber sensor[9].

To clearly understand the size dependence of strong exciton photon coupling in perovskite NW cavity, electromagnetic fields of photonic mode in perovskite NW waveguide cavities (inset of **Figure 3a-c**) is calculated by finite element method. The calculated waveguide modes are not purely transverse electric (TE) or transverse magnetic (TM). Instead, they are all behaved as TE-like, with weak longitudinal electric and magnetic components (*Supporting information*, **Note 3**). Perovskite NW can sustain a lot of waveguide modes, while the fundamental TE like mode with highest effective refractive index is selected as resonant cavity modes, which is also suggested by PL spectra. Considering that excitons are distributed throughout the entire crystal, not just at the position of maximum field, we thus calculate the effective mode volume $V_{eff}$, rather than mode volume[10] in the expression of coupling strength to quantify this average field strength in the F-P cavity. We calculate the Rabi splitting and oscillator strength of 21 different F-P cavities with cross-section dimension ranging from 0.29 to 1.33 μm and length ranging from 3.26 to 19.77 μm (**Figure 3e**, *Supporting information* **Note S5)**. When the effective mode volume is larger than 3.6 μm$^3$, oscillator strength is nearly unchanged with dimensionless average value of 0.052 per exciton. Similar result also appears in Rabi splitting, giving the average value of 267.8 meV. The result shows that in this "bulk" region, single oscillator strength remains constant, leading to a constant value of coupling strength. However, for effective mode volume smaller than 3.6 μm$^3$, oscillator strength increases significantly as effective mode volume decreases. As a result, Rabi splitting of 386.67 meV is obtained for a NW with effective mode volume of 0.89 μm$^3$, which is nearly 1.5 times of the value in "bulk"region. For a constant of total oscillator strength, the coupling strength is proportional to $\sqrt{1/V_{eff}}$ (equation S6). A fit of coupling strength is shown in **Figure 3e** as the red solid line with total oscillator strength of $1.7063 \times 10^9$ in regardless of the actual value of oscillator strength. We conclude that exciton polariton is enhanced in this "cavity" region due to the increase of oscillator strength. The oscillator strength can be enhanced by combination of several coherently oscillators result from either coherent excitation, which exists in both of bulk and micron optical cavities under laser excitation, or coherent filling of exciton wavefunction in nanocrystal[26]. Since the lateral dimensions of NW cavities are much larger than exciton Bohr radius of MAPbBr$_3$, coherent filling due to electronic quantum confinement effect can be excluded.[27] Considering the translational periodicity of single crystal which allowing the delocalize of bound excitons, it is concluded that the reduced mode volume of microcavities results in the increase of coherence filling of exciton wavefunction,[28] which then leads to significant enhancement of oscillator strength[26] and thereby coupling strength.

We studied power dependent PL spectra to investigate polariton properties influenced by carrier density. **Figure 4a** presents PL spectra under excitation of a 405 nm CW laser as a function of excitation power. Energy of polariton modes does not show significant variation below 150 μW. However, with further increasing of excitation power, polariton resonant peak exhibit slightly blue shifts. To further investigate power dependent polariton properties, PL spectra are fitted by multi-Lorentzian function, and then intensity of polaritons corresponding to five photonic modes are extracted out and plotted versus pumping power from 20 to 150 μW. As shown in **Figure 4b**, PL intensity $I_{em}$ increases super linearly with excitation density $I_{ex}^k$ with a function of $I_{em} = A I_{ex}^k$, where $A$ is a constant, $k$ is power law. The emission intensity increases superlinearly with excitation power even though power is quite low, suggesting dominant exciton-polariton scattering.[14] When polariton modes energy decreases from 2.24 eV to 2.15 eV, $k$ increases from 2.205 towards 2.374, respectively. That is, the lower energy modes are more pronounced with increasing excitation, which is attributed to the increase of photon component of polaritons at low energy side.[7] **Figure 4c** shows $E$-$k_z$ dispersion curve of lower polariton band modes at excitation power of 40 μW (olive line) and 600 μW (blue line), respectively. At low excitation intensity (40 μW), the polariton dispersion can be well fitted with $\Delta E_{LT}$ of 27 meV. However, when excitation intensity approaches 600 μW, $\Delta E_{LT}$ (c. a. 25 meV) is slightly smaller than that obtained under low excitation condition, suggesting the decrease of exciton oscillator strength. The Rabi splitting energy also show decreases from 345 meV to 338 meV. The decreasing of strong exciton-photon coupling at high excitation intensity is possible due to reduction of exciton binding energy as a result of carrier screening effect[29].

Although lasing has been widely reported in MAPbBr$_3$ NWs and nanoplatelets, deep understanding of lasing mechanism is still not adequate.[21, 30,31-32] Electron-hole plasma[33], localized[34] or bound excitons[35] have been reported to be responsible for perovskite lasing. Further, lasing properties of individual MAPbBr$_3$ NW is investigated to demonstrate whether polariton lasing could be realized considering the strong exciton-photon strength in MAPbBr$_3$ microcavity. **Figure 5a** shows power dependence PL spectrum of individual MAPbBr$_3$ NWs excited by a 400 nm femtosecond pulsed laser. When excitation fluence is small (3 to 15 μJ/cm$^2$), PL spectrum is dominated by a broad spontaneous emission peak with FWHM of about 22.2 nm. When excitation fluence increases to ~15 μJ/cm$^2$, sharp laser peaks can be resolved with FWHM of ~0.6 nm, which increase rapidly with further increasing of excitation fluence. As shown in **Figure 5b**, the integrated emission intensity versus excitation fluence shows a typical "*S*" shape which can be explained by a multimode laser theory.[35] The first region below 15 μJ/cm$^2$ is attributed to spontaneous emission; superlinear region occurs above 15 μJ/cm$^2$, indicating amplified spontaneous emission; further increase of the excitation fluence (>22.1 μJ/cm$^2$) leads to a linear region, indicating a full lasing action. The blue shift of lasing peak with increasing excitation fluence above threshold is conventionally explained as band filling or thermally induced refractive index change.[33] **Figure 5c** shows CW laser excited PL spectroscopy dominated by polariton and lasing spectrum above threshold of the same NW. The lasing peaks almost locate in the damped area of polariton energy. It means that polaritons are decoherent before they could maintain the final state population *via* relaxation (*Supporting information*, **Note S6**), leading to photonic lasing rather than polariton condensation[3]. Secondly, polariton lasing shows continuously blueshifts due to loss of strong coupling or related to the excitonic part of polaritons.[7,36] However, here lasing peaks remain nearly the same with slight blueshift above 22.1 μJ/cm$^2$ (see **Figure S7**). Therefore, the observed lasing in as-fabricated NWs belongs to photonic lasing. Polariton lasing (equivalently condensation) occurs only if stimulated scattering rate into the ground state overcomes polariton decay rate. Therefore, the realization of polariton condensation is not only dependent on large exciton photon coupling strength and sufficient polariton density, but also the efficiency of polariton relaxation into ground state. The predominant photon like polaritons are more likely to decay during the relaxation pathway before the observation of polariton lasing. This effect is often called relaxation bottleneck. There are several ways to overcome the bottleneck effect, including altering the detuning between excitons and confined photons to increase the exciton like polaritons,[37] or utilize other efficient relaxation pathway like polariton–polariton scattering under high polariton densities[7] and electron–polariton relaxation through the injection of an electron gas in the quantum well.[38] In further works, efforts would be put on optimizing NW cavity design to improve the relaxation efficiency of polariton.

**Conclusions**

In summary, strong coupling of photons and excitons in $MAPbBr_3$ perovskite Fabry-Pérot cavities have been demonstrated at room temperature. By reducing cavity volume, a transition from bulk exciton polaritons to cavity polaritons occurs, leading to the drastic increase of light-matter coupling strength, which is attributed to the increase of oscillator strength and electric-field intensity at small mode volume condition. The large coupling strength of perovskite cavity polaritons brings polariton effects, *i.e.* the observed significant reduction of group velocity is useful in slow-light devices. Condensation and polariton lasing of perovskite microcavity may be further achieved by improving the relaxation efficiency of polaritons. The observed room temperature exciton-polaritons and ease of fabrication make $MAPbBr_3$ an excellent polariton material and also broadens the application of perovskite materials in polariton-based devices filed.

**Methods**

*Characterization of perovskite NWs*: The morphology and structures of the as-grown samples were characterized using an optical microscope (Olympus BX53), AFM (Veeco Dimension 3100) in tapping mode, scanning electron microscopy (SEM, Zeiss Merlin), X-ray powder diffraction (XRD, Rigaku D/max-TTRIII, Cu Kα radiation) in the θ−θ geometry, and transmission electron microscopy (TEM, Tecnai G2 20 S-Twin; acceleration voltage, 200 kV). Absorption spectra were measured by PerkinElmer's LAMBDA 650 UV/Vis Spectrophotometer.

*Steady-State and spatial resolved Photoluminescence measurements*: The steady-state PL spectra measurements were performed using the 405 nm line of a solid state laser and the laser power was adjusted using neutral density filters. The laser was focused on the sample with a 100× objective lens (NA = 0.95). Backscattered signal was either projected onto a small charge-coupled device (CCD) for color imaging or collected through the focusing lens into an optical fiber that was coupled to a Princeton Instrument SP2500i spectrometer with a liquid nitrogen cooled CCD detector. A pin hole fixed on two dimensional stages was arranged before backscattered signal could be projected to CCD or optical fiber. Light could be collected from specific parts of the NW with a spatial resolution of ∼200 nm.

*Lasing Spectroscopy:* The 800 nm laser pulses were from a Coherent Astrella regenerative amplifier (80 fs, 1 kHz, 2.5 mJ/pulse), seeded by a Coherent Vitara-s oscillator (35 fs, 80 MHz), whereas 400 nm wavelength laser pulses were obtained with a BBO doubling crystal. The 400 nm wavelength laser pulses were used as the excitation source and focused on the sample with a 50× objective lens (NA = 0.90). All the other optical paths are identical to Steady-State PL measurements.

**Notes**

The authors declare no competing financial interest.

**Acknowledgements**

X.F.L thanks the support from the Ministry of Science and Technology (No.2016YFA0200700 and 2017YFA0205004), National Natural Science Foundation of China (No.21673054), Key Research Program of Frontier Science, CAS (No.QYZDB-SSW-SYS031). Q.Z. acknowledges the support of start-up funding from Peking University, one-thousand talent programs from Chinese government, open research fund program of the state key laboratory of low-dimensional quantum physics. Q.Z. also thanks funding support from the Ministry of Science and Technology (2017YFA0205700; 2017YFA0304600). This work is also supported by the National Natural Science Foundation of China (grant numbers 61307120, 11474187).

**Figure and captions**

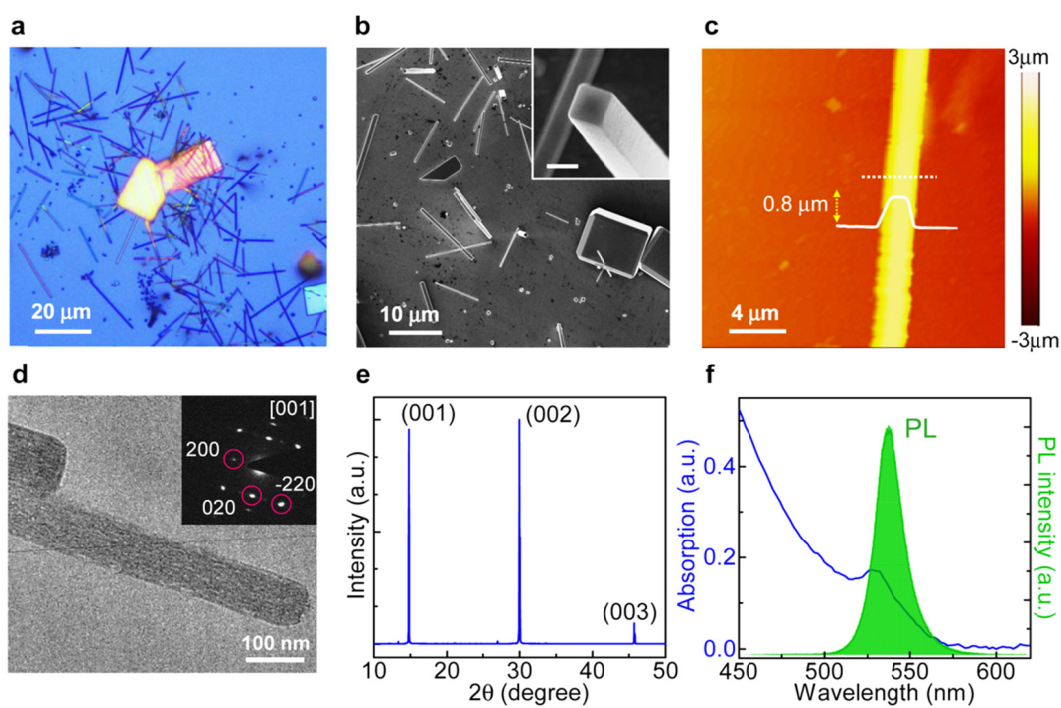

**Figure 1| Morphology and characterization of as-grown MAPbBr$_3$ nanowires. (a)** Optical image of perovskite nanowires and platelets. **(b)** Top-view SEM image of perovskite nanowires and platelets, inset is one typical image shows the rectangle section of nanowire. **(c)** AFM image of single perovskite nanowire. The white solid line represents the height profile (0.8 μm high) along the white dash line. **(d)** TEM image and SAED pattern of a single perovskite nanowire along the [001] zone axis. **(e)** XRD pattern of as-grown MAPbBr$_3$ perovskites. **(f)** Absorption spectra (blue line) and PL spectra (green area) of one typical MAPbBr$_3$ nanowire.

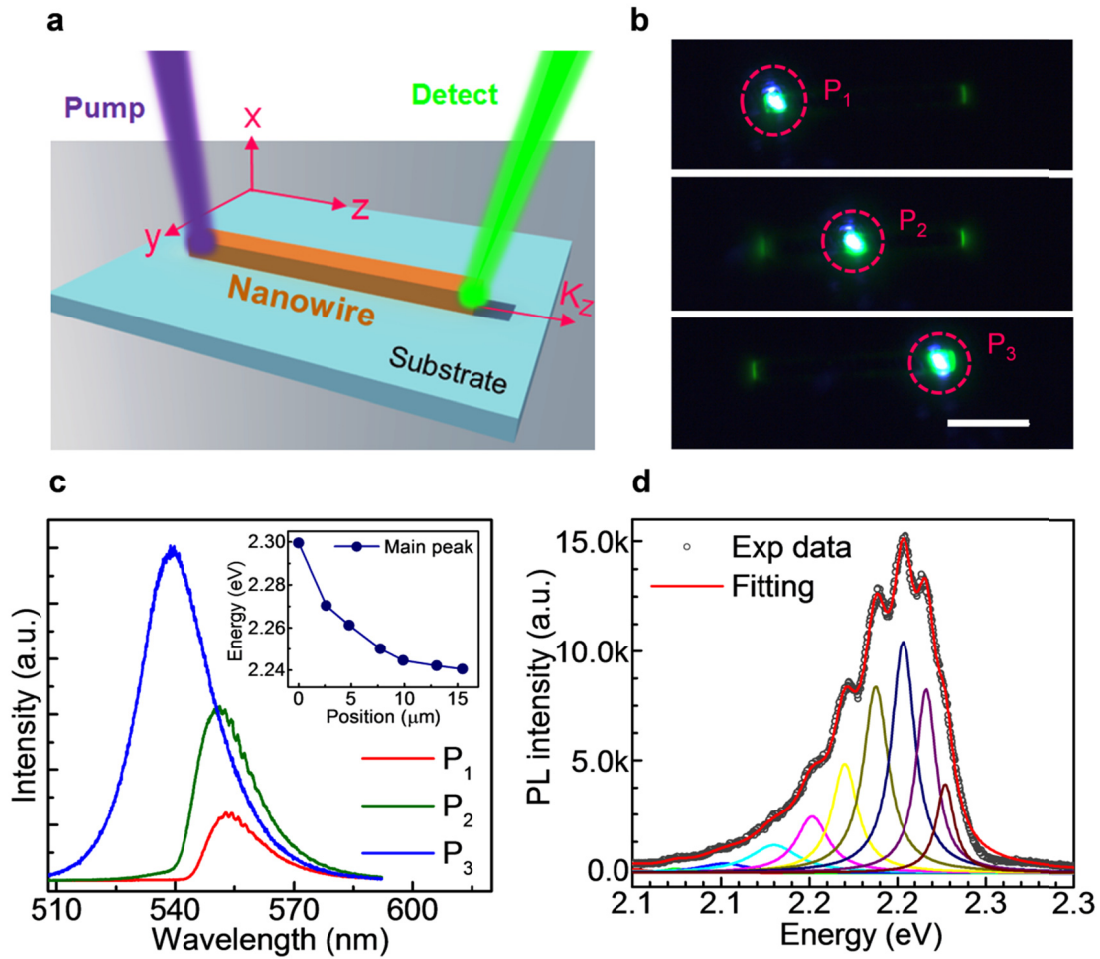

**Figure 2| Experimental geometry and spatial resolved micro-photoluminescence.**
**(a)** Schematic of experimental setup. One laser excites at one facet of nanowire and wave-guide emission collected at another facet of the same wire. **(b)** PL image for a typical MAPbBr$_3$ nanowire when excitation position moves from left end (P$_1$) to right end (P$_3$). The scale bar is 10 μm. **(c)** Corresponding PL spectra collected from the left end of MAPbBr$_3$ nanowire in figure b when excitation point at P$_1$~P$_3$ respectively. Inset is the plot of main peak position *verse* waveguide distance from excitation point to detection point. **(d)** Analysis of a typical PL spectrum collected at the right end of nanowire with left end excited with a 405 nm CW laser. The fitted curves represent different waveguide modes.

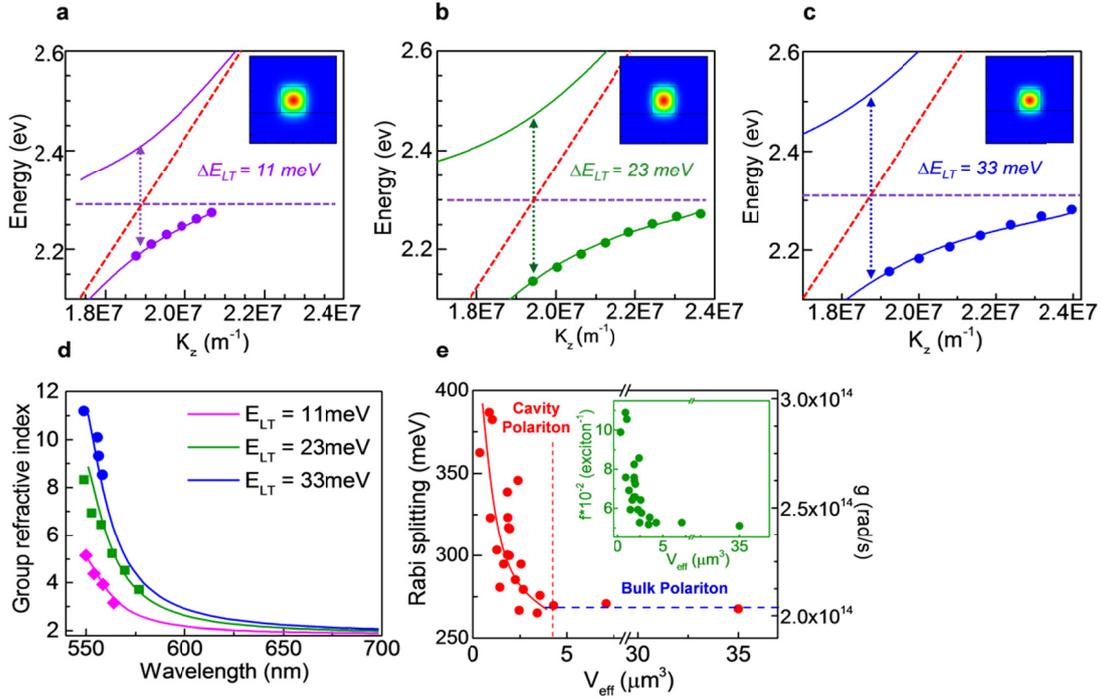

**Figure 3| Exciton polaritons of different size of MAPbBr$_3$ perovskite nanowires.**
**(a-c)** Dispersion curves of three MAPbBr$_3$ nanowires with width×length dimensions of 0.28×8.22 μm, 0.27×5.8 μm, 0.32×3.66 μm respectively. The round data points indicate Fabry-Pérot peaks, which have been placed in wave vector space at integer values of π/Lz with Lz the nanowire length. The solid lines show the results of numerical calculations for the fundamental TE like guided mode, resulting in avoided crossing at the exciton resonance (purple dash line). Red dash line, pure optical mode indicating $\hbar g$=0. As is shown by the arrows indicating 2×$\hbar g$, from **a** to **c**, accurate fits could only be obtained by increasing the longitudinal-transverse splitting (Δ$E_{LT}$). Inset is normalized electric-field distribution $|E|^2$ at the cross-section of each corresponding nanowire. **(d)** Wavelength dependence of group refractive index for nanowire with Δ$E_{LT}$ of 11 meV, 25 meV, 33 meV. The square points are experimentally determined Fabry-Pérot peaks. **(e)** Rabi splitting 2$\hbar*g$ *versus* effective mode volume for 21 nanowires with length ranging from 3.3 to 19.8 μm. For nanowires with large effective volume, the Rabi splitting remains constant (blue dash line), belonging to bulk polariton region. For smaller volumes with cavity polariton, the Rabi splitting increases by nearly 1.5 times which can be fitted by equation S6 with $n*f$=1.7063×10$^9$. Inset is corresponding exciton oscillator strength.

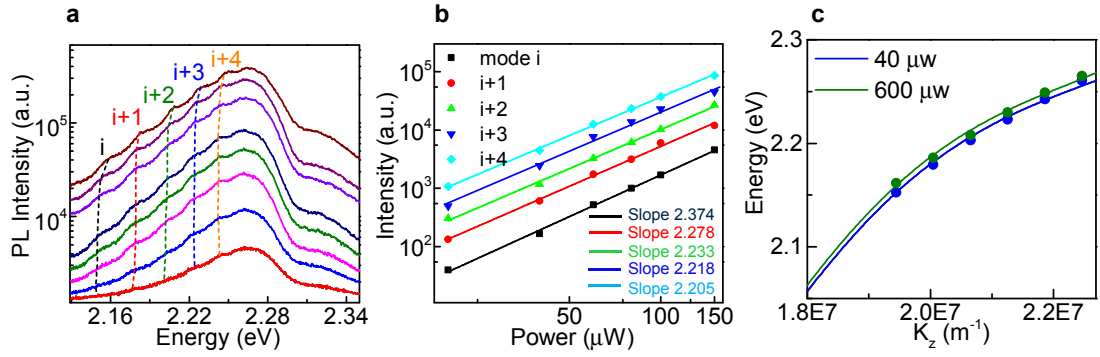

**Figure 4| Power dependence of spatial resolved PL and related dispersion property. (a)** Plots of emission spectra of a MAPbBr$_3$ nanowire in logarithmic scale at different excitation power (red, 20 μW; blue, 40 μW; magenta, 60 μW; olive, 80 μW; navy, 100 μW; violet, 150 μW; purple, 200 μW; wine, 300 μW). The dashed lines indicate the position of Fabry-Pérot emission peak maxima with the excitation intensity. **(b)** Excitation power dependence of integrated emission intensity of the fives peaks in Figure 4 (a) as a function of excitation power; the emission intensity increases superlinearly with the excitation intensity. **(c)** Dispersion curves for the modes at excitation intensities of 40 μW (blue squares) and 600 μW (green squares), which can be well fitted with polariton dispersion with $\Delta E_{LT}$ of 27 and 25 meV, respectively.

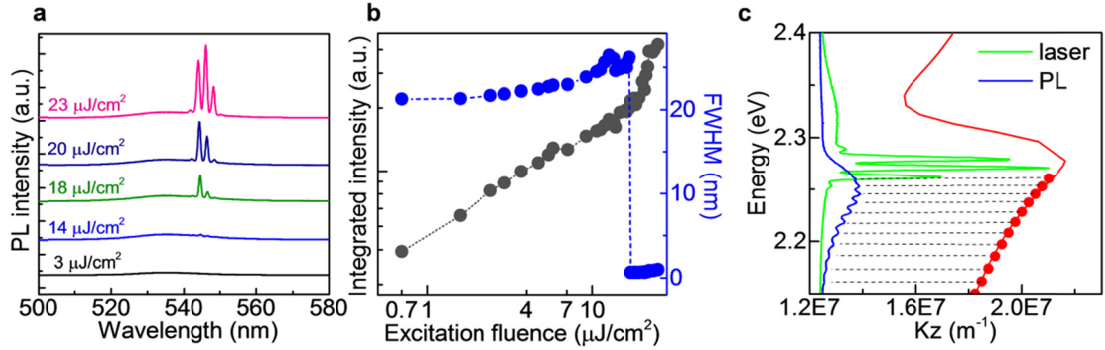

**Figure 5| Laser versus spatial resolved PL for MAPbBr$_3$ perovskite nanowire. (a)** Emission spectra of single MAPbBr$_3$ perovskite nanowire under different excition intensity, which shows the evolution from spontaneous emission to lasing. **(b)** Integrated output emission (black dots, logarithmic scale) and full width at half-maximum (blue dots) as a function of pumping fluence. The integrated output emission curve shows the threshold region as a "kink" between the two linear regions of spontaneous emission and lasing, indicating the threshold of 15 μJ/cm$^2$. **(c)** The lasing emission spectra (green line) and spatial resolved photoluminescence spectra (blue line) of identical MAPbBr$_3$ perovskite nanowire plotted in the same energy coordinate. The red dots displayed in wave vector space with integer values of $\pi / L_z$ are Fabry-Pérot peaks extracted from spatial resolved PL spectra, which can be fitted with polariton dispersion curve (red line).

**Graphic of TOC**

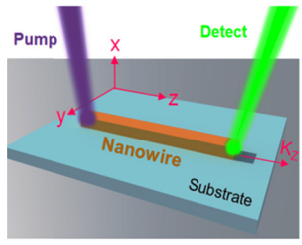 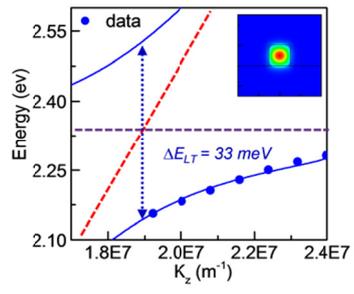 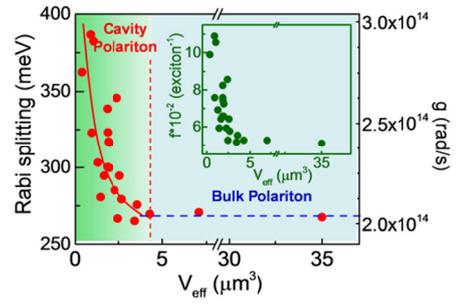